\documentclass[a4paper,fleqn,usenatbib]{mnras}

\usepackage[T1]{fontenc}
\usepackage{ae,aecompl}
\usepackage{graphicx}	
\usepackage{amsmath}	
\usepackage{amssymb}	

\def \aj {AJ}
\def \mnras {MNRAS}

\def \apj {ApJ}

\def \apjl {ApJL}
\def \aap {A\&A}
\def \nat {Nature}
\def \araa {ARAA}

\def \aaps {A\&A Suppl.}

\newcommand{\angstrom}{\mbox{\normalfont\AA}}

\def\lesssim{\mathrel{\hbox{\rlap{\hbox{\lower4pt\hbox{$\sim$}}}\hbox{$<$}}}}
\def\gtrsim{\mathrel{\hbox{\rlap{\hbox{\lower4pt\hbox{$\sim$}}}\hbox{$>$}}}}

\long\def\symbolfootnote[#1]#2{\begingroup%
\def\thefootnote{\fnsymbol{footnote}}\footnote[#1]{#2}\endgroup}

\title[Polarimetry of SN 2017egm]{{\it RINGO3} polarimetry of the Type I superluminous SN 2017egm}

\author[J.R. Maund et al.]{Justyn R. Maund$^{1}$\thanks{j.maund@sheffield.ac.uk}\thanks{Royal Society Research Fellow}, Iain Steele$^{2}$, Helen Jermak$^{2}$, J. Craig Wheeler$^{3}$\newauthor and Klaas Wiersema$^{4}$
\\
$^{1}$ Department of Physics and Astronomy, University of Sheffield, Hicks Building, Hounsfield Road, Sheffield, S3 7RH, U.K. \\
$^{2}$ Astrophysics Research Institute, Liverpool John Moores University, IC2 Liverpool Science Park, 146 Brownlow Hill,\\
	Liverpool, L3 5RF, U.K.\\
$^{3}$ Department of Astronomy, University of Texas at Austin, C1400 University Station, Austin, TX, 78712, U.S.A.\\ 
$^{4}$ Department of Physics, University of Warwick, Coventry CV4 7AL, U.K.\\
}

\date{Accepted XXX. Received YYY; in original form ZZZ}

\pubyear{2018}

\begin{document}
\label{firstpage}
\pagerange{\pageref{firstpage}--\pageref{lastpage}}

\maketitle
\begin{abstract}
The origin of the luminosity of superluminous supernovae (SLSNe) is an unresolved mystery, and a number of very different physical scenarios (including energy injection from magnetars, collision with a dense circumstellar medium and pair instability-induced explosions) have been invoked. The application of polarimetry to normal SNe has been shown to probe the three-dimensional structure of exploding stars, providing clues to the nature of the explosion mechanism.   

We report imaging linear polarimetry observations of the Type I SLSN 2017egm, in the galaxy NGC 3191, conducted with the Liverpool Telescope and the {\it RINGO3} instrument.  Observations were acquired at four epochs, spanning $4 - 19$ days after light-curve maximum, however, polarization was not detected at a level of $>3\sigma$.  At +7 and +15 days, and in the average over all epochs, we find a possible polarization signal, detected at a significance of $\approx 2\sigma$ in the ``blue" channel.  This signal is seen, primarily, in the Stokes $q$ parameter, with a corresponding polarization angle consistent with the orientation of the spiral arm in proximity to the position of SN 2017egm.  We interpret this as indicating that any polarization, if present, originates from dust in the host galaxy rather than being intrinsic to the SN itself.  Despite its apparent peculiarities, compared to other Type I SLSNe, the polarization characteristics of SN 2017egm are consistent with the previously reported low polarization of other SLSNe of this variety.
\end{abstract}


\section{Introduction}
\label{sec:intro}
With the discovery of the extremely bright supernovae (SNe) 2005ap \citep{2007ApJ...668L..99Q} and 2006gy \citep{2007ApJ...666.1116S}, it was realised that there exists a distinct class of ``superluminous" (SL) SNe, whose extreme brightness breaks the paradigm of the energetic death of massive stars powered by the collapse of an iron core.  Superluminous supernovae (SLSNe) constitute some of the brightest transients in astronomy, reaching absolute magnitudes of $M \sim -21\,\mathrm{mag}$ \citep[for a review see][]{2012Sci...337..927G}.  The current record for the brightest transient is ASASSN-15lh, which has been interpreted as being a SLSN \citep[][however, for an alternative explanation see \citealt{2016NatAs...1E...2L}]{2016Sci...351..257D}.  The origin of their luminosity still remains unclear with a number of different competing engines being invoked to explain the high luminosities, such as the decay of significant quantities of radioactive nickel produced in pair-instability induced explosions \citep{2009Natur.462..624G}, energy injection from the spin down of a newly formed magnetar \citep{2010ApJ...717..245K} and interaction with a dense circumstellar medium \citep[see e.g.][]{2007Natur.450..390W, 2012ApJ...760..154C}.

Controversially, the ``smoking gun" piece of observational evidence for the nature of the engine responsible is still absent.  The bulk of observations of SLSNe are concerned with photometry or spectroscopy  and are unable to differentiate between the different models; and similarly the  different models themselves do not necessarily provide exclusive predictions for these observables \citep{2017Natur.551..210A}.   In addition, as found by \citet{2017arXiv170801623D}, the specific interpretations of the light curve depend on when during the light curve evolution one is observing.

For the observations of Type Ia and core-collapse SNe, polarimetry has been used as a probe of the three-dimensional structures of these explosions  \citep[for a review see][]{2008ARA&A..46..433W}.   The application of polarimetry may therefore shed light on the possible energy injection mechanism for SLSNe, by being sensitive to: 1) the geometry of the ejecta (and the underlying sources of excitation); and 2) the processes by which light propagates through the ejecta.
 
\begin{figure}
\includegraphics[width=7cm,angle=270]{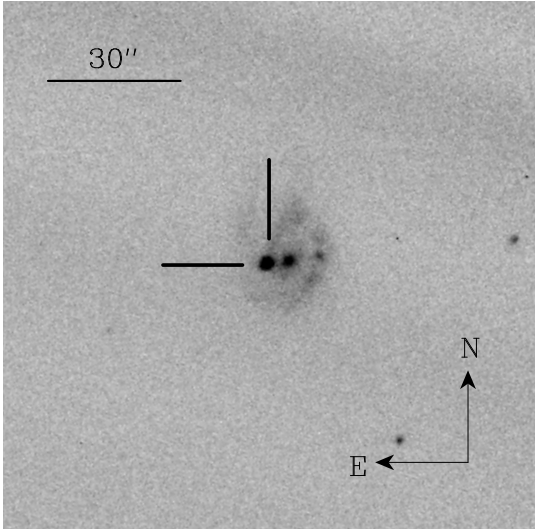}
\caption{A portion of the LT {\it RINGO3} ``green" image of SN 2017egm (indicated by the cross-hairs) in NGC~3191 on 2017 June 24.  The total field size, corresponding to the diameter of the $50\%$ vignetted image, is $246\arcsec$. }
\label{fig:image}
\end{figure}

Here we present multi-colour polarimetric observations of the superluminous Type I SN 2017egm in the galaxy NGC 3191 (see Figure \ref{fig:image}), acquired with the Liverpool Telescope (LT) {\it RINGO3} instrument.  SN 2017egm was discovered by Gaia, as Gaia 17biu, on 2017 May 23 \citep{2017TNSTR.591....1D}.  With a redshift of $z = 0.03$ for the host galaxy\footnote{Quoted from the NASA/IPAC Extragalactic Database - https://ned.ipac.caltech.edu/}, SN 2017egm is the nearest hydrogen-poor Type I SLSN yet discovered, occurring in a large spiral galaxy in close proximity to the nucleus (see Fig. \ref{fig:image}).  A number of studies, such as \citet{2017ApJ...845L...8N} and \citet{2017ApJ...851L..14W} have presented contradictory interpretations of photometric and spectroscopic observations of this event.

\section{Observations and Data Reduction}
\label{sec:obs}
SN 2017egm was observed using the Liverpool Telescope (LT) located on the Canary Island of La Palma, using the {\it RINGO3} polarimeter\footnote{http://telescope.livjm.ac.uk/TelInst/Inst/RINGO3/}\citep{2012SPIE.8446E..2JA}.  Observations were conducted at four separate epochs: 2017 June 24.9, 2017 June 27.9, 2017 July 05.9, 2017 July 08.9 (all dates are UT).  These epochs approximately correspond to $4 - 19$ days after the $g$-band lightcurve maximum \citep{2018ApJ...853...57B}.

\begin{figure}
\includegraphics[width=4.75cm,angle=270]{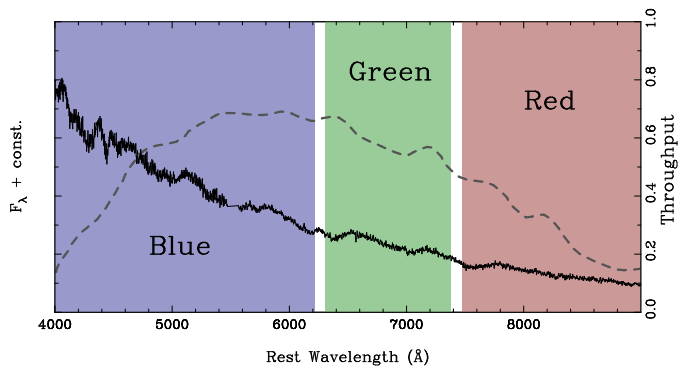}
\caption{The spectrum of SN 2017egm at +9.1 days post-maximum \citep{2018ApJ...853...57B} roughly corresponding to time period covered by our {\it RINGO3} polarimetric observations.  Overlaid are the wavelength regions corresponding to the three {\it RINGO3} channels and (dashed grey line) the instrumental throughput of {\it RINGO3} \citep{arnoldphd}.}
\label{fig:obs:spec}
\end{figure}

Each set of observations of SN~2017egm consisted of $8 \times 74.5\mathrm{s}$ exposures.  {\it RINGO3} uses a rapidly spinning polaroid to make 8 separate measurements of the  polarization components of the target source every $\sim 2.5\,\mathrm{secs}$.  A system of two dichroics is used to redirect the light to three separate channels, each with its own camera,  covering blue (``e" - $3500 - 6400\, { \angstrom}$), green (``f" - $6500 - 7600\, { \angstrom}$) and red (``d" - $7700 - 10000\, {\angstrom}$) wavelengths each.  The bandpasses for the three constituent channels of the {\it RINGO3} observations and the features in the spectrum of SN 2017egm they cover, at these epochs, are shown in Figure \ref{fig:obs:spec}.  

The {\it RINGO3} observations were bias and flat-field corrected as part of the instrument data reduction pipeline.  Photometry of the sources in the field was conducted using SExtractor \citep{1996A&AS..117..393B}, utilising an aperture of $5\,\mathrm{px}$ in diameter uniformly across all images and channels at all polaroid position angles.  The sky contribution across the aperture was derived from the background map automatically generated by SExtractor.  The initial calculation of the intensity normalised Stokes $q (= Q / I)$ and $u (= U / I)$ parameters were determined from this photometry following the methodology of \citet{2002A&A...383..360C}. To calibrate for the effects of polarization induced by the {\it RINGO3} instrument we followed prescription of \citet{2016MNRAS.458..759S}.  Observations of 324 zero and highly polarized standards, observed between 2017 May 31 and July 31, were used to calibrate for the level of instrumental polarization ($q_{0}$ and $u_{0}$), the zero angle  offset for the polarization angle and the degree of instrumental depolarization, using reference values found in \citet{1992AJ....104.1563S} and \citet{2007ASPC..364..523H}.  In the case of the polarized standards, a low order polynomial was used to convert between the degree and angle of polarization reported in the standard Johnson-Cousins filter system to the wavelength ranges appropriate for the three {\it RINGO3} channels.  The average zero angle offset ($K$) and the inverse degree of instrumental depolarization (where $D = \mathrm{``expected\, polarization"}/\mathrm{``measured\, polarization"}$) for each channel, calculated over the long baseline of observations of standards, are presented in Table \ref{tab:obs}.  Our measurements are consistent with previous values determined from zero and polarized standards over longer baselines presented by \citet{jermakphd} and \citet{arnoldphd}.  Uncertainties on the observed Stokes parameters were derived using Monte Carlo techniques.

\begin{table*}
\caption{Derived instrumental calibration parameters for {\it RINGO3} \label{tab:obs}}
\begin{tabular}{lcccc}
\hline\hline
Detector	&	$q_{0}$	&	$u_{0}$	& $K$		& $D$ \\
			&	$(\%)$	&	$(\%)$	& (degrees) &		\\
\hline
Blue (``e")	& $-0.377 \pm 0.009$ & $-1.706 \pm 0.008$ & $125.2 \pm 0.3$ & $1.05 \pm 0.02$\\
Green (``f")& $-0.910 \pm 0.015$ & $-3.239\pm 0.014$ & $125.6 \pm 3.2$ & $1.04 \pm 0.01$ \\
Red (``d")	& $-1.146\pm 0.022$ & $-3.265 \pm 0.020$ & $125.3 \pm 0.4$ & $1.02 \pm 0.02$ \\
\hline\hline
\end{tabular}
\end{table*}

\section{Results}
\label{sec:obs}
\begin{figure}
\includegraphics[width=7cm, angle=270]{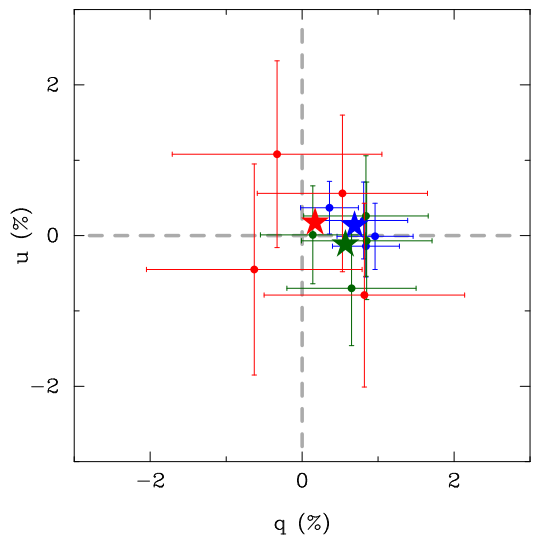}
\caption{The Stokes parameters for SN 2017egm at four separate epochs as presented in Table \ref{tab:res} ($\bullet$).  The points are colour-coded according the wavelength range of the three {\it RINGO3} channels.  The average Stokes parameters for each wavelength range, reported in Table \ref{tab:avg}, are presented by the starred points ($\star$).}
\label{fig:res:stokes}
\end{figure}

\begin{figure}
\includegraphics[width=7cm, angle=270]{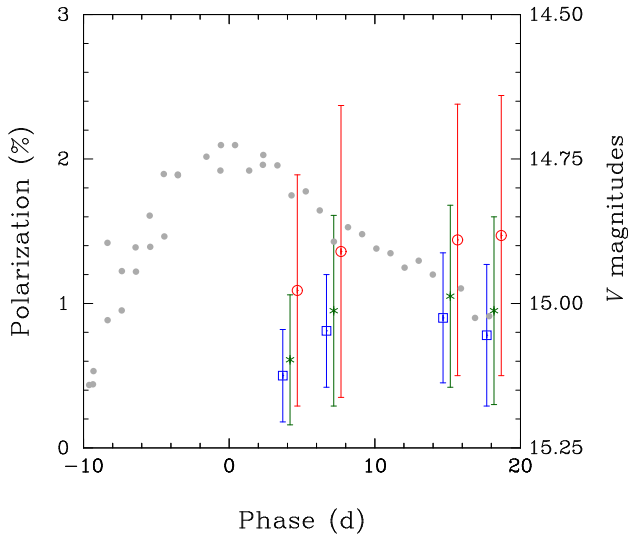}
\caption{The evolution of the degree of polarization of SN~2017egm with time.  The points indicate the polarization measured through the ``blue" ($\mathbf{\Box}$), ``green" ($\ast$) and ``red" ($\circ$) {\it RINGO3} channels.  Also shown, in grey, is the $V$-band light curve reported by \citet{2018ApJ...853...57B}}.
\label{fig:res:lc}
\end{figure}

The measured polarization properties of SN~2017egm are presented in Table \ref{tab:res} and shown on the Stokes $q-u$ plane on Figure \ref{fig:res:stokes}.  The evolution of the degree of polarization, with respect to the evolution of the $V$-band photometric light curve is shown on Figure \ref{fig:res:lc}.  For individual epochs there is no clear significant level of polarization, with the uncertainty dominated by the systematics associated with the determination of the instrumental depolarization.  There are substantially better levels of signal-to-noise associated with the blue and green channels, reflecting the blue colour of SN~2017egm at these times (see Fig. \ref{fig:obs:spec}) and the better sensitivity in these channels.  In Table \ref{tab:avg} we present the weighted-average polarization derived across all four epochs and find some possible evidence for marginal polarization at blue wavelengths, with the highest levels of signal-to-noise, which may indicate that any polarization signal, if present, is preferentially associated with the Stokes $q$ parameter.

Following the spectral evolution of SN~2017egm reported by \citet{2018ApJ...853...57B}, the intrinsic polarization in the three {\it RINGO3} bandpasses are expected to be dominated by wavelength-independent continuum polarization, with possible contamination from polarization that might be associated with lines of Fe II in the blue bandpass, Si II in the green bandpass and O I and Ca II in the red bandpass.

The data have not been corrected for the interstellar polarization (ISP) which, for broad-band imaging polarimetry, is generally difficult to determine for the full dust column to the target SN (including the host galaxy contribution). SN 2017egm, and its host galaxy, are located opposite the Galactic centre and at a relatively high Galactic latitude ($b = 54^{\circ}$), with the Galactic extinction in the direction of the SN corresponding to $E(B-V) = 0.0097 \pm 0.0005\,\mathrm{mag}$ \citep{2011ApJ...737..103S}, implying the ISP arising in the Galaxy is likely to be low (given $p_{\rm ISP} < 9 \times E(B-V) \%$ assuming a \citealt{1975ApJ...196..261S} polarization correlation, such that for SN~2017egm $p_{\rm ISP} < 0.09 \%$).  This expectation is confirmed by HD 89021 (located $\sim 3.56^{\circ}$ from the SN line of sight) in the \citet{2000AJ....119..923H} catalogue, for which the measured polarization is only $p = 0.010	\pm 0.120 \%$.  If we are to believe the slight preference for polarization signal, in the blue and green channels, being found in the Stokes $q$ parameter, i.e. $\theta \rightarrow 0^{\circ}$, this might be consistent with the apparent orientation of the spiral arm containing SN 2017egm in the galaxy NGC 3191 appearing to lie approximately North-South, with the SN located almost due East of the host galaxy nucleus (see Fig. \ref{fig:image}).  Given the expectation that the ISP arising from dust grains in other galaxies should run parallel to the magnetic field lines running along spiral arms \citep{1987MNRAS.224..299S}, the slight preference for positive Stokes $q$ may reflect an ISP contribution arising in spiral arm of NGC 3191 and the observed polarization is not intrinsic to SN 2017egm itself.

\begin{table*}
\caption{Polarimetry of SN 2017egm\label{tab:res}}
\begin{tabular}{llccccc}
\hline\hline
Date	 		& Detector 	& $q$		& $u$		& $p$		&	$\theta$  			 & $V$\\
(MJD)		&			& $(\%)$		& $(\%)$		& $(\%)$		&     (degrees) & (mag)$^{1}$\\
\hline
2017 Jun 24	&Blue 		&$0.36 (0.38)$ 	&$0.37 (0.35)$ 	&$0.50 (0.32)$ 	&$22.8 (26.0)$	 & 14.76\\
(57928.9)		&Green		&$0.14 (0.69)$	&$0.01 (0.65)$ 	&$0.61 (0.45)$ 	&$14.0 (49.3)$ 	\\
			&Red		&$0.53 (1.12)$ 	&$0.56 (1.04)$ 	&$1.09 (0.80)$ 	&$21.9 (39.0)$	\\

2017 Jun 27	& Blue		&$0.84 (0.44)$ 	&$-0.14 (0.41)$ &$0.81 (0.39)$ &$-5.3 (17.6)$ 	& 14.82\\	 
(57931.9)		& Green		&$0.84 (0.82)$ 	&$0.26 (0.80)$ 	&$0.95 (0.66)$ 	&$9.9 (32.1)$ 		\\
			& Red		&$-0.63 (1.42)$ &$-0.45 (1.40)$ &$1.36 (1.01)$ &$24.8 (62.4)$ 	\\
			
2017 Jul 5		& Blue		&$0.96 (0.50)$	&$-0.01 (0.44)$ &$0.90 (0.45)$ &$-1.1 (17.4)$	& 14.94\\
(57939.9)		& Green		&$0.65 (0.85)$ 	&$-0.70 (0.76)$ &$1.05 (0.63)$ &$-18.0 (31.8) $	\\
			& Red		&$0.82 (1.32)$ 	&$-0.79 (1.22)$ &$1.44 (0.94)$ &$-11.6 (38.4)$	\\	

2017 Jul 8		& Blue		&$0.81 (0.58)$ 	&$0.20 (0.51)$ 	&$0.78 (0.49)$ 	&$6.8 (24.7)$ 	& 15.00	\\
(57942.9)		& Green		&$0.85 (0.86)$ 	&$-0.07 (0.78)$ &$0.95 (0.65)$ &$-0.2 (33.0)$ 	\\
			& Red		&$-0.33 (1.38)$ &$1.08 (1.24)$ &$1.47 (0.97)$ 	&$41.9 (39.4)$	\\
\hline
\end{tabular}
\\
$^{1}$ Interpolated $V$-band magnitude, at the epochs of the {\it RINGO3} observations, derived from the lightcurve presented by \citet{2018ApJ...853...57B}
\end{table*}

\begin{table}
\caption{Average polarimetric properties of SN 2017egm \label{tab:avg}}
\begin{tabular}{lcccc}
\hline\hline
Detector	&	$q$			&	$u$			&	$p$		 & $\theta$\\
			&	$(\%)$		&	$(\%)$		&	$(\%)$	 & (degrees)\\
\hline
Blue			&$0.69(0.23)$	& $0.13 (0.21)$	& $0.63 (0.23)$ & $5.3(14.5)$\\
Green		&$0.57(0.40)$	& $-0.12 (0.37)$& $ 0.31 (0.40)$ & $-6.0(34.0)$ \\
Red			&$0.17(0.65)$	& $0.17(0.60)$& $0.24 (0.63)$ & $22.5(180)$\\
\hline
\end{tabular}
\end{table}

\section{Discussion \& Conclusions}
\label{sec:conc}
The lack of a significant detection of polarization, at individual epochs, associated with SN~2017egm is consistent with the generally low levels of polarization previously measured for other Type I SLSNe.  The polarization levels measured here are consistent with the limited spectropolarimetry reported by \citet{2018ApJ...853...57B} at comparable epochs from $-1$ to $+9$ days. \citeauthor{2018ApJ...853...57B} claim a significant detection at the level of $p \sim 0.43 \pm 0.09\%$ with a polarization angle of $-19 \pm 6^{\circ}$ over a wavelength range approximately corresponding to our ``red" observation. These levels are consistent with the average polarization and polarization angle measured from our $RINGO3$ observations, and confirm the preference for the polarization signal to occur in Stokes $q$.  Although \citet{2018ApJ...853...57B} claim that the lack of strong wavelength dependence to the polarization is inconsistent with a \citet{1975ApJ...196..261S} ISP law, it is important to remember that at low polarization levels the wavelength dependence may be difficult to discern \citep[see e.g. ][]{2016MNRAS.457..288R} to the point of irrelevance and this situation may be especially compounded in instances of low signal-to-noise.  Although \citet{2018ApJ...853...57B}  do identify possible evidence for modulation of the polarization across spectral features, some of these regions are associated with strong sky absorption and emission features (and may not be not intrinsic to the SN).

Our polarization measurements of SN~2017egm, associated with maximum light, are consistent with an almost spherical photosphere.  Our limits on the degree of polarization, for no ISP, constrain departures from a spherical symmetry, assuming an underlying oblate spheroidal structure, to an axial ratio of $>0.85$ \citep{1991A&A...246..481H}.  Indeed, the observation of suppression of flux at ultraviolet wavelengths, due to significant line opacity \citep{2017arXiv171101534Y}, may imply that the bluer wavelengths of our data, down to the atmospheric cutoff, could be intrinsically depolarized (similar to Type Ia SNe; see e.g. \citealt{2013MNRAS.433L..20M}), and therefore this wavelength region is most likely representative of the ISP.  Unless there were some fortuitous anti-alignment of the ISP and intrinsic polarization of the SN, our limited evidence would support an axial ratio for the SN that is close to unity.

The low levels of polarization are consistent with previously reported polarimetry of the Type I SLSNe LSQ14mo \citep{2015ApJ...815L..10L,2017ApJ...843L..17L} and SN\,2015bn \citep{2016ApJ...831...79I, 2017ApJ...837L..14L}.  For LSQ14mo, \citet{2015ApJ...815L..10L} observed similar levels of polarization, in imaging polarimetry at 5 epochs spanning -7 to +19 days relative to the light curve maximum.  \citeauthor{2015ApJ...815L..10L}  inferred an almost spherical ejecta with limits on the axial ratio for an ellipsoidal configuration of $> 0.9$.  The combined observations of SN 2015bn presented by \citet{2016ApJ...831...79I} and \citet{2017ApJ...837L..14L} covered a large time range of -24 to +27 days and -20 to +40 days, relative to the light curve maximum, respectively.  From these observations, including two epochs of spectropolarimetry, a more complex picture of the evolution of the polarization of SLSNe emerged.  Both \citeauthor{2016ApJ...831...79I} and \citeauthor{2017ApJ...837L..14L} observed a fundamental change in the polarization properties of that SN, consistent with ejecta becoming more asymmetric with depth.  Such ``phase transitions" might not be unexpected, and certainly the presence of discontinuities in the geometries of the ejecta of normal SNe have been previously witnessed through polarimetry \citep{2006Natur.440..505L,2009ApJ...705.1139M}.  As noted by \citet{2017arXiv170801623D}, significant quantities of Ni ($1 - 10 M_{\odot}$) are required to power the late-time lightcurves of Type I SLSNe.  Unless there are significant asymmetries (inconsistent with polarimetric measurements) an alternative mechanism must be responsible for the luminosity around the lightcurve maximum.   It is conceivable that if a more extended base-line of observations of SN2017egm were available a similar change in the polarization characteristics of the SN might have been observed.  In addition, \citeauthor{2016ApJ...831...79I} observed a wavelength dependence in the degree of polarization across the spectrum of SN~2015bn at the two epochs.  They claimed this could be due to an internal engine such as a magnetar or could be due to the wavelength dependence of the significant opacity due to large quantities of metals in the ejecta (possibly indicating a pair-instability event).  Such rises in polarization at redder wavelengths, due to a relative decrease in line opacity, have been previously seen in Type Ia SNe \citep{2001ApJ...556..302H,2012A&A...545A...7P}. Unfortunately, given the range of signal-to-noise in our observations in the different {\it RINGO3} channels, it is not possible to discern if there is any wavelength dependence.

A note of caution is, however, required in the comparison between the polarimetric properties of SN~2017egm and those of other Type I SLSNe.  It has been remarked that both the location, with respect to the host galaxy, and the photometric evolution of SN~2017egm are peculiar.  \citet{2017ApJ...849L...4C} determined the metallicity at the region of the SN in NGC 3191, finding solar or just above solar abundances (alternatively, \citealt{2017arXiv170803856I} derive lower metallicities and estimate an initial progenitor mass $>20M_{\odot}$).  \citet{2017ApJ...851L..14W} observed that the shape of the peak of the lightcurve was unlike those of magnetar-powered models, instead being consistent with collision of the ejecta with a dense shell of circumstellar material (CSM).  Unlike spectropolarimetry of the interacting, possible impostor SN 2009ip \citep{2014MNRAS.442.1166M,2017MNRAS.470.1491R}, the lack of strong polarization for SN~2017egm may imply that the CSM completely encloses the ejecta, providing a spherical contact surface that maximises the efficiency of the conversion of kinetic energy of the ejecta to radiative energy.  If the light curve peaks of Type I SLSNe are generally produced by CSM interaction, the published polarimetry of these objects all suggest spherical symmetry is a key ingredient.

From our observations we can exclude significant levels of polarization for SN~2017egm, around maximum light, that is commonly associated with core-collapse SNe.  The precision of our polarization measurements are limited by a small, but non-negligible degree of polarization arising from {\it RINGO3} .  Future designs for a high-time resolution polarimeter with the Liverpool Telescope Multiwavelength OPTimized Optical Polarimeter \citep[MOPTOP;][]{2016SPIE.9908E..4IJ,2017IAUS..324..357J} will use a dual-beam design, with throughput increased by a factor of $\sim 2$ compared to {\it RINGO3}, and will yield polarization precisions $<0.1\%$.  MOPTOP will provide a unique and powerful observing mode on a two-metre telescope for efficient polarimetric followup of transients in the next decade. 
\section*{Acknowledgements}

The research of JRM is supported through a Royal Society University Research Fellowship.  The research of JCW is supported in part by the Samual T. and Fern Yanagisawa Regents Professorship in Astronomy.

The Liverpool Telescope is operated on the island of La Palma by Liverpool John Moores University in the Spanish Observatorio del Roque de los Muchachos of the Instituto de Astrofisica de Canarias with financial support from the UK Science and Technology Facilities Council.

\bibliographystyle{mnras}


\bsp	
\label{lastpage}
\end{document}